\documentclass[a4paper,twoside]{article}
\newcommand{\affil}[1]{$^{\rm #1}$}

\usepackage[authoryear]{natbib}
\bibpunct{(}{)}{;}{a}{}{,}
\usepackage{graphicx}

\date{} 

\newcommand{\arcdeg}{\mbox{$^\circ$}}

\newcommand{\ex}[1]{\mbox{$\times 10^{#1}$}}

\newcommand{\Msol}{\mbox{M\raisebox{-.6ex}{$\odot$}}}

\newcommand{\kms}{\mbox{km s$^{-1}$}}
\newcommand{\Jb}{\mbox{Jy beam$^{-1}$}}

\newcommand{\lesssim}{\mbox{\raisebox{-0.3em}{$\stackrel{\textstyle <}{\sim}$}}}
\newcommand{\gtrsim}{\mbox{\raisebox{-0.3em}{$\stackrel{\textstyle >}{\sim}$}}}
\newcommand{\vapp}{\mbox{$\beta_{\rm app}$}}
\newcommand{\phn}{\phantom{1}}

\title{\large\bf\flushleft VLBI Constraints on Type I b/c Supernovae}
\author{\parbox{\textwidth}{\flushleft
\vspace{-0.5cm}
{\it Michael F. Bietenholz\affil{A,B}}\\
\vspace{0.4cm}
{\small \affil{A}\,Hartebeesthoek Radio Observatory, PO Box 443, Krugersdorp, 
1740, South Africa} \\
{\small \affil{B}\,Department of Physics and Astronomy, York University, Toronto,
M3J~1P3, Ontario, Canada}}}

\begin{document}
\twocolumn[
\begin{changemargin}{.8cm}{.5cm}
\begin{minipage}{.9\textwidth}
\vspace{-1cm}
\maketitle
\small{\bf Abstract:}
VLBI observations of supernovae and gamma-ray bursts provide
almost the only way of obtaining spatially resolved information about
the sources.  In particular, a determination of the expansion velocity
of the forward shock, as well as the geometry of the fireball and its
evolution with time are possible for relatively nearby events,
provided they are radio bright.  Monitoring the expansion of the shock
front can provide information on the density profiles of both the
circumstellar material and on the ejecta.  VLBI observations can also
potentially resolve GRB jets which are not directed along the line of
sight, providing crucial confirmation of relativistic expansion in
such objects.  This review gives an overview of recent results from
supernovae, including the Type I b/c SNe 2011dh, 2009bb, and 2007gr,
and discusses the prospects for future observations.


\medskip
\medskip
\end{minipage}
\end{changemargin}
]
\small

\section{Introduction}

Type I b and c supernova are core-collapse supernovae (SNe) that show
no hydrogen, and no hydrogen or helium, respectively, in their optical
spectra\footnote{Note that the classification may be time-dependent.
For example, SN~2008D was first classified as being of Type~Ic, but
later strong He lines emerged and it was re-classified as being of
Type Ib \citep{Modjaz+2009}.  SN~2011ei, on the other hand, appeared
to be of Type IIb early on, but had rapidly fading H lines, so that if
observed on the order of one week later it would have been classified
as Type Ib \citep{Milisavljevic+SN2011ei}.},
and refer to them collectively as ``Type I b/c''.  I will also mention
on occasion Type IIb supernovae, which are ones that show hydrogen
lines early on that subsequently disappear.  Type I b/c as well as IIb
SNe are thought to arise from the core collapse of a massive star
which has lost most or all of its envelope either due to stellar winds
or due to mass-transfer to a binary companion prior to the SN
explosion, and for this reason are sometimes known as ``stripped
envelope'' SNe \citep{Clocchiatti+1996}.  They have been of especial
interest because they have been associated with gamma-ray bursts
(GRB)\footnote{In this paper I will use the general term GRB, but am
  in fact only referring to long-duration bursts.  GRBs are generally
  classed as long when they have lengths $> 2$~s.  However, there is
  recent evidence that some bursts as short as 0.8~s also originate in
  the collapse of a massive star \citep{Piran+2013}, and thus might
  occur in conjunction with a Type I b/c SN and therefore be of
  relevance to the discussion in this paper.}.
For a review of the radio observations of GRBs, \citet{GranotvdH2013}.

In this review I will discuss in particular the results of Very Long
Baseline Interferometry (VLBI) radio observations of Type I b/c SNe.
The primary reason for VLBI observations is that the unmatched
resolution of VLBI provides almost the only way of obtaining spatially
resolved information about the sources.  VLBI can reach angular
resolutions of 0.1 milli-arc\-second, which corresponds to a distance
of $\sim$1 light-month at 50~Mpc.  VLBI observations therefore
potentially allow us to measure the size, shape, and speed of the
expanding fireballs.

Generally in supernovae, the radio emission is synchrotron emission,
which arises from the combination of magnetic fields and relativistic
particles, both of which are produced by the shocks generated by the
SN\@.  A particular advantage of radio observations in this field is
that radio emission traces the {\em fastest} ejecta,
which are difficult or impossible to see optically.  The explosive
ejection of material in a SN generally produces a forward shock where
the ejecta impact upon their surroundings, as well as a reverse shock
which is driven back into the ejecta. In the well studied case of the
Type IIb SN~1993J, \citet{SN93J-4} give arguments that the outer
bounds of the radio emission region are closely associated with this
forward shock location.

It was generally thought that the most common progenitors of Type I
b/c SNe were Wolf-Rayet stars \citep{vDykLF2003a,
  WoosleyHW2002}.  However, recently evidence has emerged that
suggests that lower mass He stars make up a substantial fraction,
perhaps the majority, of Type I b/c progenitors
\citep{Eldridge+2013,Smith+2011a}.
The progenitors of both classes of SNe are thought to be similar,
although there is evidence suggesting that the Type Ic progenitors may
be somewhat more massive than the Type Ib ones
\citep{Kuncarayakti+2013}.  However, given that to date there has been
no definitive detection of a Type I b/c progenitor, such evidence must
remain somewhat inconclusive (although note that \citealt{vDyk+2003}
and \citealt{Cao+2013} report possible detections of both type Ib and
Ic progenitors).

Type I b/c SNe can exhibit very large ejection velocities, in some
cases relativistic.  SNe where the optical spectrum shows broad
lines implying particularly large velocities ($\gtrsim 20,000$~\kms)
are often classed as ``broad lined''.  Type II SNe, by contrast, have
not generally been seen to have high ejection velocities, with the
exception of ones of Type IIb, such as SN~2003bg
\citep{Hamuy+2009} and SN~2011dh \citep{Marion+2011} which do show
high ejection velocities.
Only 5 to 10\% of all Type Ic SNe have broad lines.

Type I b/c SNe are particularly interesting because they have been
associated with long-duration GRBs.  So far, all the GRBs that have
been reliably associated with a spectroscopically confirmed SN have
been associated with one of Type I b/c.  In the cases where the
spectrum of the SN allowed a more precise classification, the SNe were
of Type Ic with broad lines.  Although many, perhaps all GRBs show a
supernova-like ``bump'' in their lightcurves, only in relatively
nearby examples can the accompanying SN be identified more
conclusively in the spectrum \citep{Bersier2012}.
Note however, that the great majority of Type I b/c SNe are not
accompanied by GRBs.  In fact, even of the subset of broad-lined Type
Ic SNe, only $\sim$20\% are associated with a GRB event.  

\sloppypar The most popular model for long duration GRBs is the
so-called ``collapsar'' model, \citep{Woosley1993, MacFadyenWH2001},
in which a core-collapse supernova produces or triggers a central
engine (accreting, rapidly spinning compact object).  The central
engine drives a highly relativistic jet, while the more spherical SN
explosion is powered by neutrinos.  An observable GRB is produced when
the jet is oriented close to the line of sight, and the strong Doppler
boosting causes the observed strong gamma-ray emission.  Such events
also produce emission at lower frequencies, called the afterglow.  In
particular, radio afterglows are often produced.  The radio emission
is produced later in the evolution of a GRB, once the velocities have
become sub-relativistic and the emission isotropized.  Unlike the
gamma-ray emission, therefore, the radio emission is not strongly
beamed.

So it seems that both the stripped-envelope nature of the supernova
and the presence of particularly high ejection velocities are
required, but not sufficient for a SN to produce a GRB\@. The
tentative picture that emerges is that a small subset of Type I b/c
SNe are characterized by relativistic ejection velocities, and
thus have broad-lined optical spectra.  Some of these produce bright,
cosmological (``normal'') GRBs which emit $> 10^{49.5}$~erg in gamma
rays, while more of them produce low-luminosity bursts like
GRB~0908425, which emit $< 10^{48.5}$~erg in gamma rays. 
In the case of the bright GRBs, the ejection is highly collimated
(opening angles of $\sim$0.1 radian), and is probably due to
a relatively long-lived (at least tens of seconds) jet that emerges from
the collapsing star.  This jet occurs in conjunction with a more
isotropic SN event, but in the case of an observed GRB where
the jet is presumably oriented near our line of sight, the emission
produced by the jet is usually brighter than that produced by the
supernova.

\sloppypar In a bright GRB, the high-energy emission is thought to be
produced by shocks internal to the jet.  The emission at longer
wavelengths, termed the ``afterglow'', and in particular the radio
emission, is thought to be produced by the external shocks as the jet
interacts with the surrounding medium.

In the low-luminosity GRBs the ejection is likely less collimated
(opening angle $\gtrsim$1 radian), and the high energy emission is
probably associated with the supernova shock breakout
\citep[e.g.][]{Kulkarni+1998, NakarS2012}, and the jet, if present,
does not emerge from the stellar surface, or is relatively weak
compared to the more isotropic supernova shock breakout.  Note that
probably the majority of GRBs are of the low-luminosity variety, and
only a minority of GRBs are ``normal'', but the latter are far more
easily detected \citep{Liang+2007}.  \citet{Soderberg+2006c} showed
that the volumetric rate of low-luminosity GRBs is comparable to that
of broad-lined Type I b/c SNe.

At present, it is not well understood why these high ejection
velocities occur in some SNe, and what causes the collimation.
Clearly information as to the size, shape, and speed of the
expanding ejecta, such as can be obtained with VLBI, can provide
important observational constraints in attempting to unravel these
mysteries.

However, SNe which are sufficiently bright to be observable with VLBI
occur only rarely.  Only a fraction of SNe are ever detected in the
radio, and of those, only for ones that are relatively nearby is VLBI
imaging useful.
For example, after 90 days, a SN expanding with a normal
non-relativistic speed of $20,000$ \kms\ could be resolved with 22-GHz
VLBI (resolution $\sim$0.2 mas with a global array) out to a distance
of $\sim$5~Mpc, while one expanding with an apparent speed of $c$
could be resolved out to $\sim$80~Mpc.  Given a model of the
morphology of the radio emission, geometric models can be fit directly
to data in the visibility (Fourier-transform) plane, and angular sizes
determined with accuracies which depend on the signal-to-noise, but
can be up to 20$\times$ higher than the nominal imaging
resolution. Such angular size estimates, however, are dependent on the
geometrical model assumed for the emission, which is not well known, in
particular for relativistic events.  This model uncertainty will
usually dominate the uncertainty in the size estimate.
Nonetheless, in most cases angular size estimates can be made with a
fractional accuracy of better than 50\%, leading to estimates of the
apparent expansion velocity of similar accuracy provided the distance
and explosion date are well constrained. \citep[see][for more
detailed discussions of this process]{SN2008D-VLBI,SN93J-2}.

In the case of relativistic ejecta, it is possible that the apparent
expansion speed becomes superluminal.  Indeed, in the case of GRB
030329, associated with SN~2003dh, VLBI observations showing that the
radio emitting region was expanding with an apparent speed of 3 to 5
$c$ during the first few months after the GRB
\citep{Taylor+2004,Taylor+2005, Pihlstrom+2007} constitute the
most compelling evidence for relativistic expansion in GRB events.

However, given the small fraction of SNe that are sufficiently radio bright
that such imaging is possible and near enough that it is useful,
the number that has so far been observed is small.  Furthermore, since
the radio emission from Type I b/c SNe usually fades over a period of
months\footnote{Some Type II SNe remain radio bright for many
  years, with notable examples being SN~1993J, SN~1986J and SN~1979C
  \citep[see, e.g.][]{Bartel2009}, but Type I b/c SNe usually
  fade quickly.}
so the period over which any one SN can be observed will be
limited.  In this review I will summarize the constraints that have
been so far obtained through such observations.

Note that broadband radio observations of SNe which cannot be
resolved by VLBI provide a second, important albeit less direct
constraint on the size of the forward shock region.  The broadband
spectral energy distributions of SNe typically are synchrotron
self-absorbed below some turnover frequency which is usually in the
range of radio observation (0.1 to 100 GHz).  For a given radio
luminosity, the frequency at which synchrotron self-absorption becomes
important depends only on the size of the emitting region, and
therefore the size can be estimated if the spectral peak can be
identified (provided the distance is known).  This method is typically
useful for measuring the size of the radio emission region early on in
the evolution of the SN (see for example
\citealt{ChevalierF2006}).  It is, however, dependent on several
assumptions, including that of equipartition between relativistic
electrons and magnetic field in the post-shock region.  In the case of
the Type IIb SN~2011dh, an estimate of size and expansion of the
forward shock was obtained in this fashion, with the estimates
subsequently confirmed by VLBI measurements
\citep{Bietenholz+SN2011dh-III}.

Table \ref{tsne} gives an overview of the Type I b/c SNe which have
been observed with VLBI to date.  VLBI observations were also obtained
of two further SNe which were discovered in the radio, and are thus of
uncertain type: SN~2008iz in M82, which was likely of Type~II, but
could have been of type I b/c also \citep{Brunthaler+2010a}, and the
unusual SN~1986J \citep[see, e.g.][]{SN86J-2}, which is generally
classed as Type~IIn, but for which a possible type Ib classification
was also suggested \citep{Leibundgut+1991}.
As can be seen, only a small number of SNe have been observed with
VLBI, and in the remainder of this paper, I will discuss most of these
relatively rare events in more detail.

\begin{table*}[th]
\begin{center}
\caption{Type I b/c supernovae observed with VLBI}
\begin{tabular}{lll@{}c@{}c@{}l}
\hline Supernova & Type & Galaxy & Distance & $L_{\rm peak, 8.4 GHz}^a$ & References\\[5pt]
                 &      &        & {\footnotesize(Mpc)}  & \footnotesize{($\times 10^{27}$~erg~s$^{-1}$ Hz$^{-1}$)}  \\
\hline 
SN 1994i  & Ic      & M51      & \phn9 & \phn2 & \parbox[t]{170pt}{\small Bietenholz \& Bartel, unpublished} \\
SN 2001em & I b/c   & NGC 7142 & 80    & 12  & \parbox[t]{170pt}{\small \citet{SN2001em-1,SN2001em-2}, 
                                               \citet{Paragi+2005}, \citet{Schinzel+2009}} \\
 SN 2003L  & I c    & NGC 3506 & 92    & 30 & {\small \citet{Soderberg+2005}}\\
 SN 2003gk & I b/c  & NGC 7460 & 45    & $<0.5$ & {\small \citet{SN2003gk-VLBI}} \\
 SN 2007gr & I b/c  & NGC 1058 & 10    & 0.1  & \parbox[t]{170pt}{\small \citet{Paragi+2007}, \citet{SN2007gr-Soderberg}} \\
 SN 2007uy & I b    & NGC 2770 & 27    & 1.1 & \parbox[t]{170pt}{\small \citet{vdHorst+2011}} \\
 SN 2008D  & I b/c BL$^b$ 
                    & NGC 2770 & 27    & 2.5 & \parbox[t]{170pt}{\small\sloppypar \citet{SN2008D-VLBI}, \citet{vdHorst+2011}} \\
 SN 2009bb & I b/c BL$^b$ 
                    & NGC 3278 & 40    &  50    & {\small \citet{SN2009bb-VLBI}} \\
\hline
\end{tabular}
\\[5pt]
\end{center}
$^a$ The peak spectral luminosity of the supernova at 8.4~GHz \\
$^b$ ''BL'' indicates a supernova classified as broad-lined
due to the presence of broad absorption lines in the early optical
spectrum.\\ 
\label{tsne}
\end{table*}

\section{SN~2001em and other ``Off Axis'' Afterglows}
\label{s2001em}

A consequence of the idea that GRBs are produced by highly collimated
jets oriented near the line of sight is that for each observed GRB,
there should be numerous similar events but with the jets not oriented
near the line of sight.  These events, which I will call ``off-axis''
events, and which are also known as orphan afterglows, would not
produce observable gamma-rays.  However, since the radio emission
mostly arises once the source becomes mildly relativistic so that its
radio emission is more consequently isotropic, the off-axis events
should not be significantly more difficult to detect than on-axis ones
in the radio.  Radio wavelengths were shown to be good for detecting
such off-axis events \citep[e.g.][]{Paczynski2001, GranotL2003}, and
in a constant density medium, an off-axis jet is expected to peak
between 1 and 3~yr after the event \citep{vEertenZM2010, GranotL2003}.
Note, however, that more recent results show that for a
wind-stratified medium (with density $\propto r^{-2}$), and using a
reasonable range for parameters such as the energy of the explosion,
the circumstellar density and the efficiency of magnetic field
generation and particle acceleration at the shock, a wide variety of
lightcurves can be produced, many of which are much fainter than the
canonical models, and some of which overlap with the expected radio
lightcurves of non-relativistic SNe \citep{SN2003gk-VLBI}.  It is
therefore likely that only a small fraction of such off-axis
relativistic jets will be observably bright in the radio.

\sloppypar Nonetheless, it was suggested that SN~2001em, at a distance
of $\sim$80~Mpc, and spectroscopically classified as a Type I b/c
supernova, might possibly harbour such an off-axis GRB jet on the
basis of its unusually high X-ray and radio luminosities, with the
latter reaching a maximum only $\sim$1000~d after the explosion
\citep[][see also Paczy\'nski
  2001]{GranotR2004}\nocite{Paczynski2001}.  Several groups obtained
VLBI observations of SN~2001em \citep{Stockdale+2005, SN2001em-1,
  Paragi+2005}.
The most constraining of these results was that of \citet{SN2001em-1},
obtained at $t \sim 3.2$~yr, where here and hereafter I will use $t$ to
indicate the time after the time of the shock breakout.  The VLBI
observations were not consistent with the relativistic expansion
expected of a GRB jet, but rather suggested an average expansion
velocity of $20,000_{-12,000}^{+7000}$ \kms, typical of what is found
in ordinary non-relativistic SNe.

Subsequent VLBI observations \citep{SN2001em-2, Schinzel+2009} have
further confirmed the non-relativistic expansion, with the most recent
$3\sigma$ limit on the expansion speed being 9000~\kms, and have also
placed non-relativistic limits on the proper motion
\citep{Schinzel+2009}.  Furthermore, broad H$\alpha$ lines have
appeared in SN 2001em's optical spectrum, prompting a
re-classification to Type IIn \citep{SoderbergGK2004}.  An alternate
model for SN~2001em, not involving a GRB event or any relativistic
ejection, was proposed by \citet{ChugaiC2006}, which involves the
interaction of normal, non-relativistic, SN ejecta with a
massive and dense circumstellar shell, produced by mass loss of the
progenitor, and which seems to be able to account for the
observational data.

So far, no off-axis events have been detected in blind surverys
\citep[e.g.][]{Levinson+2002, Gal-Yam+2006}.  Note, however, that
blind surveys of a sensitivity sufficient to ensure detection of
off-axis events at the predicted brightness levels are still
prohibitively expensive in observing time.
The generation of wide-field radio instruments currently coming on
line or envisioned for the near future, such as ASKAP, APERTIF, LOFAR
and MeerKAT may well detect off-axis GRBs in blind searches
\citep[e.g.][]{Murphy+2013}.

\sloppypar However, since GRBs arise in conjunction with supernovae,
of which many are detected optically, the strategy of looking for
off-axis GRB events accompanying known Type I b/c SNe suggested
itself.  So far, two different searches\footnote{Note that
\citet{Berger+2003} also searched for radio emission
from Type I b/c SNe, however, only a few of their observations
were at late times $t > 1$~yr, and in none of those cases
was radio emission detected.} 
for late-time radio emission from Type I b/c SN did not yield any
unambiguous detections: \citet{Soderberg+2006b} carried out a search
for late-time radio emission from Type I b/c SNe, observed 68 SNe at
late times with the VLA\@.  A further set of radio observations of 59
Type I b/c SNe was carried out by \citet{SN2003gk-VLBI}, who also did
not detect any late-time radio emission due to off-axis GRB jets, and
concluded that fewer than $<2$\% of all Type I b/c SNe, and $<30$\% of
the broad-lined ones, are associated with jet comparable to those seen
in bright GRB afterglows.  However, they also showed that for more
realistic parameters for the circumstellar density as well as the
efficiency of particle acceleration and field generation, the radio
emission from off-axis jets could be several orders of magnitude lower
than previously predicted, and that therefore only a small minority of
off axis jets would likely produce observable radio emission, and that
consequently the non-detection of such radio emission so far is
readily explained even if a large fraction of SNe Ib/c do harbor
relativistic jets with energies comparable to that of GRB jets.

In Bietenholz et al.'s survey for late-time radio emission, only a
single supernova, SN~2003gk, at $\sim$45~Mpc, was reliably detected,
with an 8.4-GHz spectral luminosity of $\sim$6\ex{27} erg s$^{-1}$ at
$t \simeq 6$~yr.
However, VLBI follow-up observations showed that SN~2003gk's average
(over $t \simeq 8$~yr) expansion speed was $\sim$10,000~\kms, ruling
out a relativistic jet, and suggesting that also in this case, 
interaction with the a dense shell in the circumstellar medium was
powering the radio emission \citep{SN2003gk-VLBI}.

A further supernova, SN~2007bg, was also suggested to possibly harbour
an off-axis jet on the basis of its radio emission
\citep{PrietoWS2009}.  No VLBI observations of it were undertaken due
to its relatively large distance of 152~Mpc, however, also in this
case, the radio (and X-ray) emission is due to interaction of
non-relativistic ejecta with dense CSM resulting from episodic
mass-loss from the progenitor rather than to any relativistic ejection
(\citealt{Salas+2013}, see also \citealt{Soderberg2009}).

\section{SN~2008D: Jet or Shock Breakout?}
\label{s2008d}

SN 2008D was first discovered as the X-ray flash (XRF) 080109 with the
accompanying supernova being discovered soon after in the optical
\citep{SN2008D-Nature}.  No gamma-ray emission was seen even though
the source was in the field-of-view of the Burst Alert Telescope. It
is at a distance of $\sim$28~Mpc, and was, as mentioned above,
originally classified as of Type Ic with broad lines, but subsequently
re-classified as Type Ib when narrow He lines appeared
\citep[e.g.][]{Modjaz+2009}.

The origin of the X-ray flash is the subject of some debate.  Some
authors suggested that it is of supernova-shock origin
\citep{SN2008D-Nature, ChevalierF2008}, while others suggested that it
was caused by a mildly relativistic jet which penetrated the envelope
of the progenitor \citep{Mazzali+2008, Li2008b, Xu+2008}.

\begin{figure*}[th]
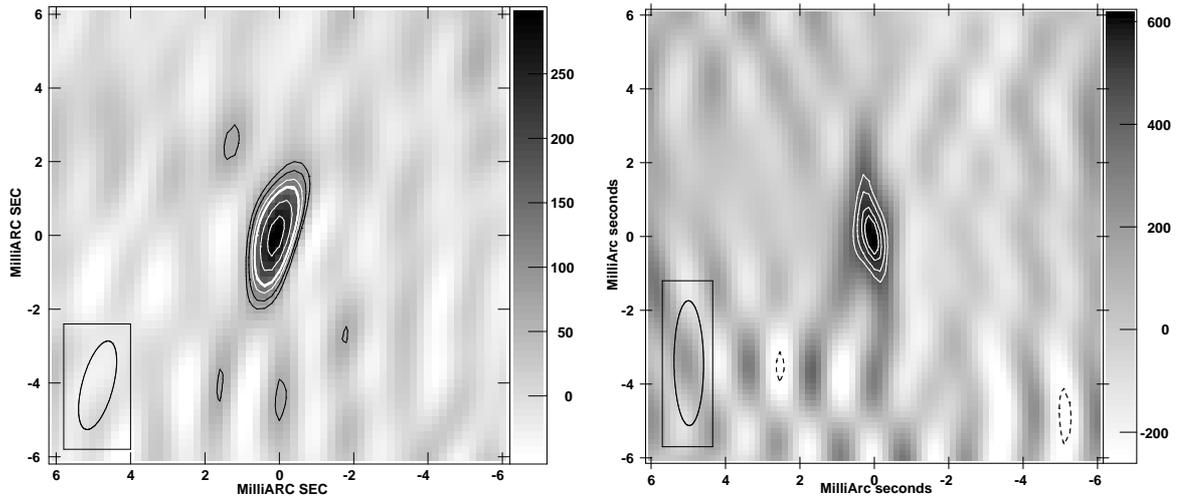

\begin{center}
\includegraphics[width=0.48\textwidth, angle=0]{fig1_sn2008D-Cband.eps}
\includegraphics[width=0.48\textwidth, angle=0]{sn09bb-rgbb1.5.eps}
\caption{VLBI images of SN~2008D and SN 2009bb.  For both images, the
  lowest contour is drawn at $3\times$ the image rms background
  brightness, and the greyscale is labelled in $\mu$\Jb, the FWHM of
  the convolving beam is indicated at lower left and north is up and
  East is to the left.
{\bf Left:} an image of SN 2008D at 5 GHz on 2008 May 21.  The
contours are drawn at $-21$, 21, 30, 40, {\bf 50}, 70, and 90\% of the
peak brightness of $296 \, \mu$\Jb, with the 50\% contour being
emphasized \citep[from][]{SN2008D-VLBI}.
{\bf Right:} an image of SN~2009bb at 8.4~GHz 2009 June 11\@. The
contours are at $-57$, 57 70,80,90 and 95\% of the peak brightness of
613~$\mu$\Jb\ \citep[see][for details]{SN2009bb-VLBI}.  I do not
consider the extensions to the south of the supernova (visible in the
greyscale, but below the $3\sigma$ contour) to be real.}
\label{fvlbiimg}
\end{center}
\end{figure*}

Radio emission was detected from SN~2008 shortly after shock breakout.
Several sets of VLBI observations were obtained.  VLBI Observations
were obtained with various arrays including telescopes of the NRAO
Very Long Baseline Array and the European VLBI Network (EVN) as well
as Arecibo, at frequencies of 22, 8.4 and 5.0~GHz, and at epochs 
$t = 28, 30, 69$ and 133~d \citep{SN2008D-VLBI, vdHorst+2011}.
Although the source was not definitively resolved at any of these
epochs or frequencies, $3\sigma$ upper limits on the apparent
expansion velocity, \vapp, of $0.71\,c$ could be set\footnote{The
quoted limit was for an assumed circular morphology, for an elongated
morphology, the $3\sigma$ upper limit on \vapp\ was $2.6\,c$
\citep{SN2008D-VLBI}.} from the observations at $t = 133$~d
\citep{SN2008D-VLBI}.  I show a VLBI image of SN~2008D in
Figure~\ref{fvlbiimg} (left).  \citet{vdHorst+2011} obtained similar
limits on the \vapp. These limits on \vapp\ ruled out a long-lived,
highly relativistic outflow, but are still compatible with a
rapidly-decelerating, mildly relativistic jet.  Recent simulations by
\citet{Bersten+2013} suggest that SN 2008D's bolometric lightcurve
cannot easily be reproduced without the presence of $\sim$0.01~\Msol\
of ~$^{56}$Ni-rich material in the outer layers of the ejecta, which
they attribute to the action of such a jet.

\section{SN~2009bb: Relativistic Ejecta?}
\label{s2009bb}

\sloppypar The nearby SN for which there was the best evidence of
relativistic expansion was SN~2009bb, detected by the Chilean
Automatic SN Search Program \citep[CHASE,][]{Pignata+2009a}, at a
distance of $\sim$40~Mpc in the nearby spiral galaxy NGC~3278,
The shock breakout date was unusually well constrained to be March
$19\pm1$~UT\@.  SN~2009bb showed broad lines, implying a photospheric
velocity of $\sim$25,000~\kms, had a moderate X-ray luminosity, and
was quickly found to be radio bright \citep{SN2009bb_Nature}.  Its
peak 8.4-GHz spectral luminosity was $\sim$5\ex{28} erg s$^{-1}$
Hz$^{-1}$, larger than that observed for any other SN I~b/c at a
similar time after shock breakout although only slightly higher than
that of SN~1998bw.

Angular sizes derived from fitting a synchrotron self-absorption model
to the broadband radio SED suggested a mean shock velocities of $0.85
\pm 0.02\, c$, assuming equipartition of energy between electrons and
magnetic fields \citep{SN2009bb_Nature}.

VLBI observations were undertaken at $t = 85$~d.  Unfortunately,
the declination of SN~2009bb was $-40$\arcdeg, and the paucity of VLBI
telescopes in the Southern hemisphere generally renders VLBI
observations difficult, and furthermore the South African telescope at
Hartebeesthoek, which provides the long baselines crucial for high
resolution, was out of commission due to a bearing failure.  The
observations were obtained at a frequency of 8.4~GHz, when the total
flux density of the SN at that frequency was 2.5~mJy, and I show the
VLBI image in Figure~\ref{fvlbiimg} (right).  Only an upper limit on
the angular size could be determined, which was 0.64~mas ($3\sigma$;
corresponding to a shock velocity of $< 1.74 c$; see
\citealt{SN2009bb-VLBI} for details).  The VLBI observations are
therefore consistent with, but do not require, moderately relativistic
expansion for SN~2009bb.

\citet{ChakrabortiR2011} interpreted SN~2009bb as having a
baryon-loaded relativistic blastwave, launched by a central-engine
driven explosion, and predicted that it would be resolvable by VLBI by
$t \simeq 200$~d.  However, by that time the flux density was already
too low ($\lesssim 1$~mJy at 8.4~GHz) to warrant further VLBI
observations.

In the case of another nearby Ic supernova, SN~2007gr, 
5-GHz VLBI observations at $t \simeq 25$ and 85~d
\citep{SN2007gr-Nature} were first interpreted to suggest relativistic
expansion.  Relativistic expansion was unexpected since the radio
luminosity of SN~2007gr was low, with the 8.4-GHz spectral luminosity
peaking at $\sim 10^{26}$~erg~s$^{-1}$~Hz$^{-1}$, almost 1000 times
lower than that of SN~1998bw.  A large angular size and therefore
large expansion velocity was suggested because the VLBI observations
recovered only a fraction of the total flux density measured at WSRT,
suggesting that a significant fraction of the flux density was on
scales larger than those measured by VLBI, and therefore setting a
lower limit on the size of the source.  However, a re-examination by
\citet{SN2007gr-Soderberg} showed that prolonged relativistic
expansion was not likely in this case.  In particular the difference
between the total flux density as measured with the WSRT and that
observed even at the shortest VLBI baselines, if interpreted as being
due to a resolved source, suggested a source much larger than expected
even for credible relativistic expansion.  \citet{SN2007gr-Soderberg}
suggested a more plausible explanation of a somewhat overestimated
WSRT total flux density and moderate coherence loss in the VLBI
observations.  \citet{Maeda2013b} also find that the radio
lightcurves of SN~2007gr suggest non-relativistic expansion.
Nonetheless, a relativstic ejection which is rapidly ($<1$ week)
decelerated is still compatible with, although by no means required by, the
VLBI measurements.

A further supernova, SN~2003L, (92~Mpc), although not showing any
broad absorption lines, had a very high radio luminosity, comparable
to that of SN~1998bw, which suggested that there might be relativistic
ejecta present.  It was observed with the NRAO VLBA at $t = 65$~d and
the results suggested an expansion velocity of only $0.2\,c$
\citep{Soderberg+2005}.

\section{Conclusions}
\label{sconc}

Type I b/c SN are particularly interesting because they can have
relativistic ejecta and because they have been associated with GRBs.
However, the nature of the relativistic ejection and the
relationship between Type I b/c SNe GRBs is far from clear.  Among the
many open questions are: What causes some SNe to have a relativistic
shock breakout?  What causes some SNe to have a comparatively
long-lived relativistic jet which can produce a GRB? Where are the
off-axis GRB jets? What is the relationship between GRBs and less
extreme supernova?  {\em VLBI observations are virtually the only way
  to obtain spatially resolved information about the explosions, and
  thus directly constrain the geometry and expansion speeds of the
  ejection.}  Unfortunately, SNe that are close and radio-bright
enough to be accessible to VLBI are rare, to date only 8 Type I b/c
SNe have been observed with VLBI (see Table~\ref{tsne}).

In none of this small sample observed so far have the VLBI
observations produced conclusive evidence of relativistic expansion,
or allowed the shape of the emission region to be determined. The VLBI
observations have, however, ruled out long-lived relativistic jets
such as are thought to produce GRBs in several cases (SN~2001em,
SN~2003gk).  The case of short-lived, more isotropic relativistic
ejection is more difficult.  Since the events are short-lived, only in
the case of very nearby events does the fireball expand to a size
resolvable by VLBI before it becomes too faint to observe.  The
evidence in these cases is consequently less conclusive, with the VLBI
observations obtained so far being compatible with, but not requiring,
the mildly-relativistic ejecta suggested by observations at other
wavelengths in the cases of SN~2008D and SN~2009bb.

Given the importance of direct observational determination of the
size, speed and shape of the expanding shock front, VLBI observations
of any SNe that are sufficiently near and radio-bright should be
undertaken.  With the increasing sensitivity due to both increasing
bandwidth, such as the 2 Gbit~s$^{1}$ bandwidth now available at the
NRAO VLBA\footnote{\tt
  https://science.nrao.edu/facilities/vlba/docs/\-manuals/oss2013a}
(to be upgraded to 4 Gbit~s$^{-1}$ in future), and the incorporation of new and
refurbished telescopes into global VLBI networks (see
\citealt{vLangevelde2013} and \citealt{Gaylard2013}), individual SNe
can be observed for longer times, and fainter ones are becoming
accessible to VLBI observations.
Although the resolution at any particular frequency is limited by the
length of the baselines and therefore limited by the Earth's diameter for
ground-based telescopes, higher sensitivity also allows observations
at higher frequencies with a concomitant increase in angular
resolution.  Higher resolution could also be obtained by using
space-based VLBI antennas such as RadioAstron \citep{Kardashev+2013},
however, due to the limited sensitivity of RadioAstron only
exceptionally radio bright SNe could be studied in this manner.
The development of e-VLBI (see e.g.\ \citealt{vLangevelde2009b})
offers quick response times, which could be crucial for resolving
relatively nearby events early on in their evolution, with the first
e-VLBI observations of a SN being those of the Type I b/c SN~2001em of
\citet{Paragi+2005}.  The availability of a VLBI array which operates
full time, such as the NRAO VLBA, rather than only in sessions during
part of the year, is also of importance due to the often short-lived
nature of these events.  It will be important also to follow up SNe
not detected in the optical, since the examples of SN~2008iz
($\sim$4~Mpc) and SN~1986J ($\sim$10~Mpc) show that even relatively
nearby SNe can go undetected in optical observations.

\section*{Acknowledgements} 
The research at York University was supported by the National Sciences
and Engineering Research Council of Canada.  I have made use of NASA's
Astrophysics Data System Bibliographic Services.  I thank N. Bartel,
J. Granot, Z. Paragi and A. Gal-Yam for useful comments on the
manuscript.

\newcommand{\araa}{Ann.\ Rev.\ Astron.\ Astrophys.}
\newcommand{\aap}{Astron.\ Astrophys.}
\newcommand{\aapr}{Astron.\ Astrophys.\ Rev.}
\newcommand{\aaps}{Astron.\ Astrophys.\ Suppl.\ Ser.}
\newcommand{\aj}{AJ}
\newcommand{\apj}{ApJ}
\newcommand{\apjl}{ApJL}
\newcommand{\apjs}{ApJS}
\newcommand{\apss}{ApSS}
\newcommand{\baas}{BAAS}
\newcommand{\memras}{Mem.\ R. Astron.\ Soc.}
\newcommand{\memsai}{Mem.\ Soc.\ Astron.\ Ital.}
\newcommand{\mnras}{MNRAS}
\newcommand{\iaucirc}{IAU Circ.}
\newcommand{\jrasc}{J.\ R.\ Astron.\ Soc.\ Can.}
\newcommand{\nat}{Nat}
\newcommand{\pasa}{PASA}
\newcommand{\pasj}{PASJ}
\newcommand{\pasp}{PASP}

\bibliographystyle{apj}
\bibliography{mybib1,snIbc-temp.bib}

\end{document}